\documentclass[fleqn,twoside]{article}
\usepackage{espcrc2}
\usepackage{epsfig}

\newcommand{\vm}{V_{m\bar{m}}(R)}
\newcommand{\beq}{\begin{equation}}
\newcommand{\eeq}{\end{equation}}
\newcommand{\beqn}{\begin{eqnarray}}
\newcommand{\eeqn}{\end{eqnarray}}
\newcommand{\bea}[1]{\beq\begin{array}{#1}}
\newcommand{\eea}{\end{array}\eeq}

\newcommand{\dual}[1]{{}^{*}{#1}}

\newcommand{\NP}[3]{{Nucl. Phys. }{\bf #1} (#2) #3}
\newcommand{\NPPS}[3]{{Nucl. Phys. Proc. Suppl.}{\bf #1} (#2) #3}
\newcommand{\PL}[3]{{Phys. Lett. }{\bf #1} (#2) #3}
\newcommand{\PRL}[3]{{Phys. Rev. Lett. }{\bf #1} (#2) #3}
\newcommand{\PRep}[3]{{Phys. Rep. }{\bf #1} (#2) #3}
\newcommand{\PR}[3]{{Phys. Rev. }{\bf #1} (#2) #3}

\newcommand{\IJMP}[3]{{Int. J. Mod. Phys. }{\bf #1} (#2) #3}

\title{
\vspace{-18mm}
\thispagestyle{empty}
\rightline{\small KANAZAWA-01-13~~~~~}
\rightline{\small ITEP-LAT/2001-03~~~~~}
\rightline{\small 15 October, 2001~~~~~}
\vspace{2mm}
Heavy monopole potential in gluodynamics
}

\author{M.~N.~Chernodub\address{ITEP, B.Cheremushkinskaya 25, Moscow, 
117259, Russia}${}^{{\mathrm{,}}}$\address{Institute for Theoretical 
Physics,  Kanazawa  University, Kanazawa 920-1192, Japan}\thanks{M. N. Ch. 
is supported by JSPS Fellowship P01023.},
	F.~V.~Gubarev${}^{\mathrm{a}}$,
	M.~I.~Polikarpov${}^{\mathrm{a}}$\thanks{M.I.P 
is partially supported by grants RFBR 00-15-96786, RFBR 
01-02-17456, INTAS 00-00111 and CRDF award RP1-2103.},
        V. I. Zakharov\address{Max-Planck Institut f\"ur Physik,
F\"ohringer Ring 6, 80805 M\"unchen, Germany}
}       

\begin{document}

\begin{abstract}
We discuss predictions for the interaction energy of the fundamental
monopoles in gluodynamics introduced via the 't~Hooft loop. At short
distances, the heavy monopole potential is calculable from first principles.
At larger distances, we apply the Abelian dominance models. We discuss the
measurements which would be crucial to distinguish between various models.
Non-zero temperatures are also considered. Our predictions are in
qualitative agreement with the existing lattice data.
We discuss further measurements which would be crucial to check the model.
\end{abstract}

\maketitle

\section{HEAVY MONOPOLE}

The fundamental monopoles can be introduced via the 't~Hooft loop
\cite{'tHooft:1978hy} on the lattice and
correspond to point-like objects in the continuum limit. On the
lattice  the monopoles are identified with the
end-points of the Dirac strings which in turn are defined as
piercing negative plaquettes.
In more detail, consider the standard Wilson
action of $SU(2)$ lattice gauge theory:
\beq
S_{lat}(U) = - \beta\sum\limits_p \; \frac{1}{2}\mathrm{Tr}\;
U_p\,.
\label{plaquette}
\eeq
Then the 't~Hooft loop is formulated (see, e.g. 
\cite{Hoelbling:2000su,forcrand} and references therein) in terms of a 
modified action $S(\beta,- \beta)$:
\beqn
S(\beta,- \beta) =
- \beta \sum\limits_{p\notin \dual \Sigma_j} \!\!
\frac{1}{2}\mathrm{ Tr}\;U_p
+ \beta\sum\limits_{p\in \dual \Sigma_j}\!\!\frac{1}{2}\mathrm{ Tr}\;U_p\,,
\nonumber
\eeqn
where $\dual \Sigma_j$ is a manifold which is dual to a surface spanned on 
the monopole world-line $j$. Introducing the corresponding partition 
function, $Z(\beta,-\beta)$ and considering a time-like planar rectangular 
$T\times R$, $T\gg R$ contour $j$ one can define
\beq
V_{m\bar{m}}(R) ~\equiv ~ -\frac{1}{T} 
\ln{ Z(\beta,-\beta)\over Z(\beta,\beta)}\,.
\label{energy}
\eeq

Since the external monopoles become point-like particles in the continuum
limit the potential $V_{m\bar{m}}(R)$ is the same fundamental quantity as,
say, the heavy-quark potential $V_{Q\bar{Q}}$ related to the expectation
value of the Wilson loop. By analogy, we will call the quantity
$V_{m\bar{m}}(R)$ the heavy monopole potential.
First direct measurements of $V_{m\bar{m}}(R)$ on the lattice were reported
\cite{Hoelbling:2000su,forcrand}. Motivated by these measurements we review
the theoretical predictions~\cite{gubarev1,gubarev2,gubarev3} on the heavy
monopole potential.

\section{POTENTIAL AT ${\mathbf{T=0}}$}

Consider first the potential $\vm$ at short distances. Then the interaction 
between the monopoles is Coulomb-like (see \cite{gubarev1,gubarev2}),
\beq
V^{Coul}_{m\bar{m}}(R) = - {\Biggl(\frac{2 \pi}{g(R)}\Biggr)}^2 \!
\frac{1}{4 \, \pi \, R} = - {\pi\over g^2(R)} \;{1\over R}\,,
\label{small-R-potential}
\eeq
where $g(R)$ is the running coupling of the gluodynamics.  
This equation makes
manifest that monopoles in gluodynamics unify Abelian and non-Abelian 
features. Namely, the overall
coefficient, $\pi/g^2$ is the same
as in the Abelian theory with fundamental electric charge $e=g$ while the running
of the coupling $g^2$ is governed by the non-Abelian interactions.

As far as the physics of short distances is concerned, the next step is to 
consider power corrections to (\ref{small-R-potential}). Theoretically, the 
prediction is that there is no linear in $R$ term at small~$R$:
\beqn
V_{m\bar{m}}(R) \!\approx \! V^{Coul}_{m\bar{m}}(R)
\! + a_0\Lambda_{QCD} \! + a_2\Lambda_{QCD}^3 R^2\dots\,,
\nonumber
\eeqn
where $a_{0,2}$ are constants sensitive to the physics in the infrared.
One can show~\cite{gubarev1,gubarev2} that the absence of the linear in $R$ 
term at short distances follows from general principles and 
eventually is related to the fact that the monopoles are not confined.

It is a common point that at large distances $\vm$  is of the Yukawa type,
\beq
\label{yukawa}
V_{m\bar{m}}(R) ~=~ - C \cdot {{e^{- \mu R}} \over R}\,,
\eeq
see, e.g., \cite{Hoelbling:2000su,forcrand,gubarev2,gubarev3,samuel,stack}.
At a closer look, however, there is  a variety of model dependent predictions for
the parameters $C$ and $\mu$. 

Historically, the first prediction for $\mu$ was obtained in \cite{samuel}.  
Namely, it was shown that to any order in the {\it strong coupling
expansion} the mass $\mu$ coincides with the mass of $0^{++}$ glueball, $\mu
= m_G$. As for the constant $C$, there is no reason for it to be the same
for the Yukawa and Coulomb like potentials ($i.e.$, $C\neq\pi/g^2$).

A different prediction arises within effective field theories with
monopole condensation. Note that the monopoles which condense are of
course {\it not} the fundamental monopoles which are introduced via the
't~Hooft loop as external probes. To describe the condensation 
one introduces a new (effective) field $\phi$
interacting minimally with the dual gluon, $B$, (for review and
references see, e.g., \cite{gubarev2,baker}). Within this model,
\beq
\label{predictionhiggs}
\mu~=~m_V\,,\qquad C ~=~ \pi \slash g^2\,,
\eeq
where $m_V$ is the mass of the vector field $B$ acquired through the Higgs
mechanism. 

It is worth emphasizing that $m_V$ can well be below the lightest glueball
mass, $m_V<m_G$. Indeed, because of the presence of the Dirac string there
is no dispersion representation for the 't Hooft loop granted. Observation
of $m_V<m_G$ via the lattice measurements would be am amusing demonstration
of existence of the Euclidean ``masses'' which have no direct counterpart in
the Minkowsi space.

\section{POTENTIAL ${\mathbf{T \neq 0}}$}
\label{Tneq0}

The screening mechanism at high temperatures is the Debye screening.  Using
the fact that classical limit of the state created by 't~Hooft loop is an
Abelian monopole pair~\cite{gubarev2,gubarev3} and utilizing the Abelian
dominance conjecture, one can estimate the mass $\mu(T)$ as $\mu^2 =m^2_D =
16\pi^2 \rho \slash e^2_3$, where $\rho$ is the density of the Abelian
monopoles and $e_3(T)$ is the three-dimensional coupling constant
corresponding to the dimensionally reduced model. This formula is valid at
asymptotically high temperatures which ensure that the monopoles form low
density gas and the dimensional reduction works well. Below we used the
results obtained in the Maximal Abelian projection~\cite{AbProj}.

To estimate the temperature dependence of $m_D$ we use the numerical results
of Ref.~\cite{Bornyakov}, where the density of Abelian monopoles was
obtained, $\rho = 2^{-7} (1 \pm 0.02) \, e^6_3$. At the tree level one can
express 3D coupling constant $e_3$ in terms of 4D Yang--Mills coupling $g$,
$e^2_3 (T) = g^2(\Lambda,T) \, T$, where $g(\Lambda,T)$ is the running
coupling calculated at the scale $T$ and $\Lambda$ is a dimensional constant
which can be determined from lattice simulations.

At present, the lattice measurements of the $\Lambda$ parameter are
ambiguous and depend on the quantity which is used to determine it. We use
two "extreme" values: $\Lambda = 0.262(18)\, T_c$  
obtained in Ref.~\cite{Heller} and~\cite{Bali} $\Lambda = 0.076(13) \, T_c$.

\section{COMPARISON WITH DATA}

Here we summarize briefly
the comparison of the lattice measurements with predictions above.

{\it Short distances.} The Coulomb-like potential (\ref{small-R-potential})
is confirmed in the numerical simulations in the classical approximation
\cite{Hoelbling:2000su,gubarev1}. There is no running of $g^2$ on this level
of course. As for the full quantum simulations the normalization
(\ref{small-R-potential}) of the potential at short distances is confirmed
within a factor of about 2, Ref.~\cite{forcrand-1}. As for the power
corrections, all the data so far \cite{Hoelbling:2000su,forcrand} are fitted
smoothly with a Yukawa potential (\ref{yukawa}).

\begin{figure}[!htb]
    \epsfxsize=7.4cm \epsffile{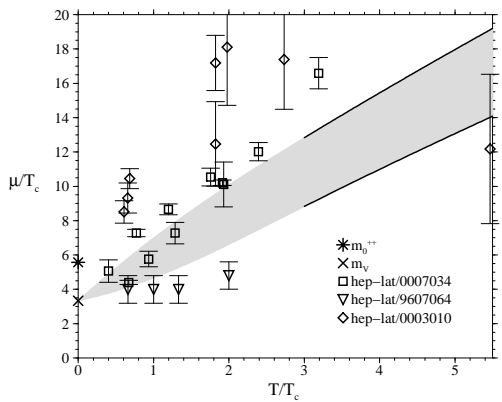}
\vspace{-12mm}
\caption{Temperature dependence of the screening mass in the heavy
monopole potential (\ref{yukawa}).}
\label{masses}
\end{figure}

{\it Screening mass at $T=0$.} In Figure~\ref{masses} the dual gluon
mass~\cite{ilgenfritz}, $m_V \approx 1 GeV$, and the mass of the lightest
$0^{++}$ glueball~\cite{Teper} are shown by cross and star respectively.
Existing data \cite{Hoelbling:2000su,forcrand} seem to be not accurate
enough to distinguish between the glueball exchange and monopole
condensation models. Moreover, no checks of the vector-particle exchange
have been made.

{\it Temperature dependence of the screening mass.} In Figure~\ref{masses}
we have summarized the existing data on temperature dependence of the
screening mass together with our predictions, Section~\ref{Tneq0}. The
direct measurement of the screening mass \cite{Hoelbling:2000su,forcrand} is
shown by the diamonds and squares, respectively. The 
theoretically expected temperature dependence of the screening mass
(described in the previous Section) is denoted by the shaded region.
Although numerically the screening mass seems to be systematically higher
than is predicted, the prediction based on the Abelian dominance hypothesis
is rather close to the numerical results within the errors.

\vspace{-7mm}
\vspace{-5mm}

The triangles in Figure~1 denote the results of Ref.~\cite{stack} obtained
in the Maximal Abelian projection using an effective Abelian monopole
action. At small temperatures these results are quantitatively in agreement
with Refs.~\cite{Hoelbling:2000su,forcrand} while at higher temperatures the
screening mass obtained in Ref.~\cite{stack} falls essentially lower. As is
explained in Ref.~\cite{gubarev3}, the prediction of Ref.~\cite{stack} is
justified at very low and very high temperatures only.

To summarize, the heavy monopole potential at small distances is of the
Coulomb type with a known overall normalization. The large--distance
potential is of the Yukawa type. According to the Abelian dominance model,
the screening mass is the same as entering the effective Ginzburg-Landau
Lagrangian relevant to the confinement. The temperature dependence of the
screening mass can also be predicted. The existing lattice data do not
contradict the predictions but allow only for a qualitative tests of the
model. Further measurements are desirable for quantitative tests.


\end{document}